\documentclass[prb,superscriptaddress,a4paper,twocolumn,showpacs,preprintnumbers]{revtex4-1}

\usepackage{color,subfigure,amsmath}
\usepackage{bm}
\usepackage{epsfig}
\usepackage{natbib}
\usepackage{graphicx}
\usepackage{amssymb,amsmath,amsbsy,amsgen,amsfonts}
\usepackage{dcolumn}
\usepackage{amsthm}
\usepackage{mathrsfs}
\usepackage{latexsym}
\usepackage{array}
\usepackage{amstext}
\allowdisplaybreaks[1]

\newcommand{\ket}[1]{\left\vert{#1}\right\rangle}
\newcommand{\ketbra}[2]{|#1\rangle \langle#2|}
\newcommand{\be}{\begin{equation}}
\newcommand{\ee}{\end{equation}}
\newcommand{\ba}{\begin{array}}
\newcommand{\ea}{\end{array}}
\newcommand{\bqa}{\begin{eqnarray}}
\newcommand{\eqa}{\end{eqnarray}}

\begin{document}

\title{Tradeoff between Leakage and Dephasing Errors in the Fluxonium Qubit}
\author{David A. Herrera-Mart\'i}
\email{dahm@nus.edu.sg}
\affiliation{Controlled Quantum Dynamics Theory, Level 12 EEE, Imperial College London, London SW7 2AZ, United Kingdom}
\affiliation{Centre for Quantum Technologies, National University of Singapore, Block S15, 3 Science Drive 2, Singapore 117543}

\author{Ahsan Nazir}
\email{a.nazir@imperial.ac.uk}
\affiliation{Controlled Quantum Dynamics Theory, Level 12 EEE, Imperial College London, London SW7 2AZ, United Kingdom}

\author{Sean D. Barret\dag}

\affiliation{Controlled Quantum Dynamics Theory, Level 12 EEE, Imperial College London, London SW7 2AZ, United Kingdom}

\date{30 August 2013}

\begin{abstract}
We present a tradeoff between anharmonicity (related to leakage) 
and pure dephasing errors 
for the fluxonium circuit. We show that in the insulating regime, i.e., when the persistent current flowing across the circuit is suppressed, the pure dephasing rate induced by a Markovian environment decreases exponentially as the impedance of the circuit is increased. In contrast to this exponential decrease, the qubit remains sufficiently anharmonic so that gate times can still be short, allowing for significant reduction in the computational error rates. 
A transition from the insulating to superconducting phases 
establishes an upper bound on the Josephson energy below which this tradeoff exists. 
\end{abstract}

\maketitle


\section{Introduction}

Superconducting circuits are a promising technology for building scalable quantum processors in the solid-state.~\cite{DiVi00b} They feature a tunable anharmonic spectrum that can be used as a two-level system, defining a qubit. Single- and two-qubit gates have been demonstrated \cite{yamamoto2003demonstration, steffen2006measurement, plantenberg2007demonstration} and the lithographic fabrication process is a mature technique, allowing for large scale production. However, as a general rule, solid state qubits couple much more strongly to their surrounding environments than systems such as atoms or photons, 
leading to far shorter decoherence times. Considerable effort has thus been invested in fighting the causes of decoherence in different superconducting circuits, \cite{vion2002manipulating,bladh2005single,koch2007charge,bylander2011noise,kim2011decoupling,kermanprl} in order to attempt to bring achievable error rates close to fault-tolerant thresholds.
Importantly, recent experimental advances point to pure dephasing (i.e.~dephasing without population relaxation) as a significant 
source of decoherence in superconducting qubits coupled to a cavity. \cite{manucharyan2009fluxonium, paik2011observation, rigetti2012superconducting} In particular, in cases where relaxation processes can be greatly reduced, \cite{paik2011observation, rigetti2012superconducting, manucharyan2012evidence} pure dephasing can lead to decoherence ($T_2$) times that are significantly shorter than the lifetime limit of $2T_1$. 

Of course, this need not be a problem if quantum gates can be performed sufficiently quickly, so that the error rate per gate is low. Given a dephasing rate $\Gamma_\varphi$, and in the absence of relaxation processes, the dephasing error probability per gate is $p_\varphi \propto \Gamma_\varphi \tau$, where $\tau \propto \Omega^{-1}$ is the time taken to perform a gate, and $\Omega$ is the Rabi frequency associated to coupling to the driving field.  
It is thus crucial for low error-rate operation that the ratio of $\Gamma_\varphi$ to $\Omega$ can be made small. One way to achieve this would be to drive the system strongly, so that gates are performed extremely quickly on the timescale set by $\Gamma_\varphi$. However, this is ultimately limited by the anharmonicity $\delta$ offered by the system, i.e., the energy difference between the first and the second transition energies, $\delta = (\Delta_{21}-\Delta_{10})$. A driving field resonant with the transition $\ket{0} \leftrightarrow \ket{1}$ will induce leakage into level $\ket{2}$ with probability
 \be 
 p_L \approx \left(\hbar\Omega/\delta\right)^2.
 \label{leakage}
 \ee
This probability has to remain well below the threshold for loss errors \cite{varnava2006loss, stace2009thresholds} to guarantee an error-free computation at the logical level. Hence, a strategy for suppressing the system pure dephasing rate without significantly compromising its anharmonicity is attractive, as it would allow the error probabilities due to both pure dephasing and leakage to be made small simultaneously.

\begin{figure}[!t]
\centering
\subfigure{
\includegraphics[scale=.25]{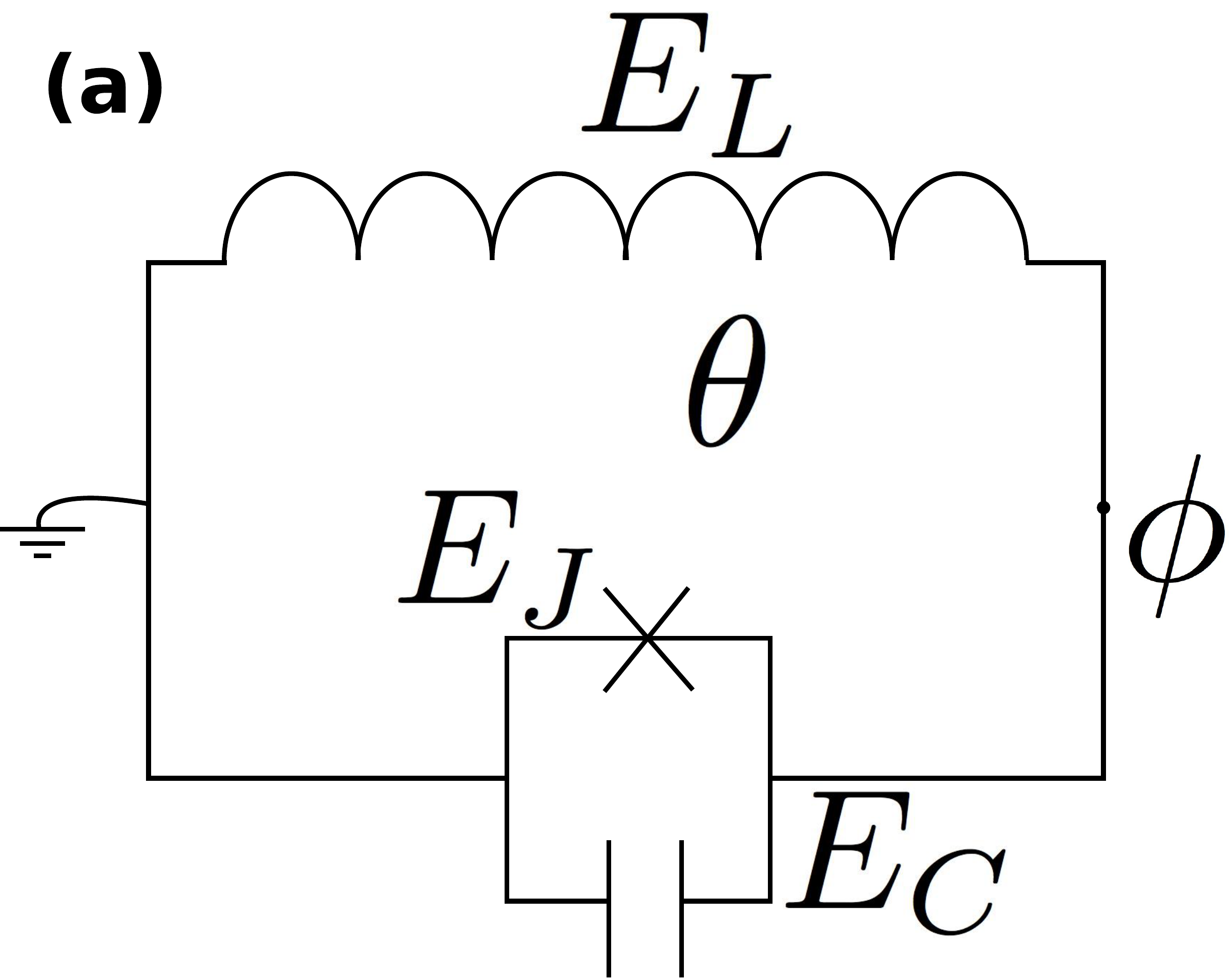}}
\subfigure{
\includegraphics[scale=0.15]{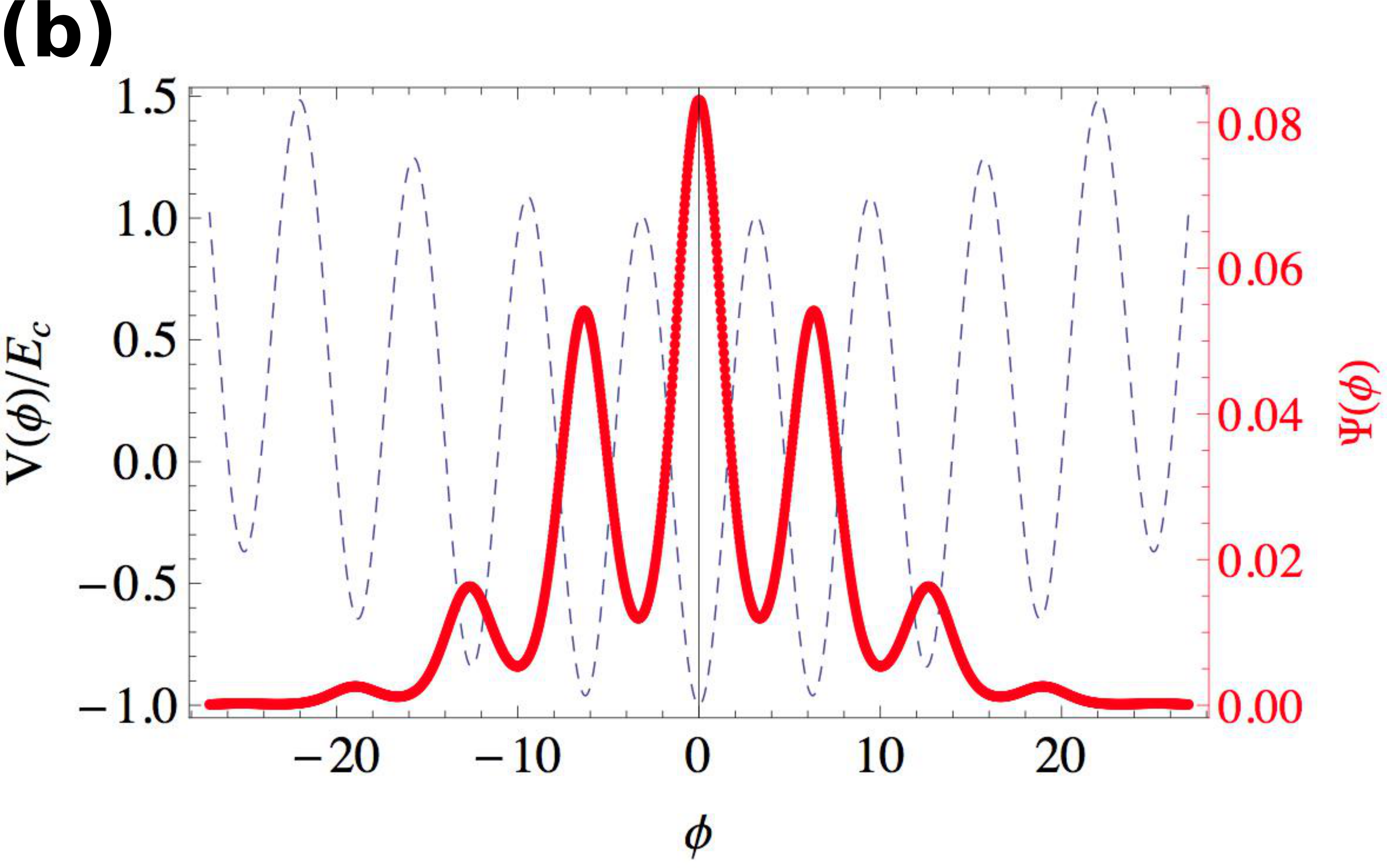}}
\caption{{(a)} Fluxonium circuit diagram. 
The gate charge $n_g$ is not depicted since the energy spectrum does not depend on it, but rather on its temporal variation. {(b)} Potential $V(\phi) = E_L\phi^2 - E_J\cos \phi$ (dashed) and ground state wavefunction (solid) for the values $E_J/E_C=1$ and $E_L/E_C=10^{-4}$.}
\label{fig1}
\end{figure}

Here, we show that by appropriate choice of operating regime 
it is possible to {\emph{exponentially}} increase pure dephasing times in a superconducting circuit (induced by coupling to a memoryless, Markovian environment), 
while reducing its anharmonicity only 
at a smaller, algebraic rate. We thus obtain a tradeoff between anharmonicity, related to leakage probability, and computational errors caused by pure dephasing. 
In Ref.~\onlinecite{koch2007charge} a similar issue was considered for the transmon circuit, where it was found 
that operating the charge qubit at large values of the Josephson energy provides protection against charge fluctuations, without compromising the qubit controllability. 
In contrast, we address this point for the fluxonium circuit, \cite{koch2009charging} which is a particularly interesting superconducting circuit exhibiting robustness against low frequency variations of both offset charge and external flux. Fig.~\ref{fig1}(a) shows a simplified version of the fluxonium circuit, 
which is operated in the regime 
$E_L/E_C \ll E_J/E_C \approx 1$, where $E_L$, $E_C$, and $E_J$ are the inductive, capacitive, and Josephson energies, respectively.

\section{The Fluxonium qubit}	  
	  
Closing a superconducting island with an inductor breaks the confinement of charge carriers, and as a consequence the conjugate charge in the island is not necessarily an integer number of Cooper pairs that have tunneled in or out. 
Equivalently, since periodicity of the superconducting phase implies that the conjugate charge must be discrete, in this case the superconducting phase $\phi$ is not periodic. Protection against low frequency charge noise is obtained for the fluxonium qubit in the same way as in other persistent current circuits, namely, by shunting the island with an inductive element in such a way that offset charge can be accounted for by choosing a different gauge. \cite{koch2009charging} Also, whenever the persistent current flowing across the circuit is suppressed, its dependence on external flux fluctuations is highly reduced.

Together with an enhanced protection against low frequency noise, the fluxonium circuit also features a suppression of quantum fluctuations of charge. As we shall shortly see, quantum fluctuations of flux are proportional to the circuit's impedance $Z_0 = \frac{\sqrt{E_C/E_L}R_Q}{2\pi}$, where $R_Q = h/(2e)^2$ is the resistance quantum. In the ground state, charge fluctuations will be inversely proportional to flux fluctuations, so in order to keep fluctuations well below one Cooper pair, the impedance must be made much larger than the resistance quantum $R_Q$. \cite{masluk2012microwave} In order to attain this, elements with large $L$ need to be created without increasing the associated capacitive energy of the inductive element, since the phase-slip amplitude decreases exponentially in the ratio $\sqrt{E_J/E_C}$ [see Eq.~(\ref{qps_amplitude})]. This has already been achieved in some experiments using chains of Josephson junctions. \cite{manucharyan2012evidence, bell2012superinductor, masluk2012microwave} Of special interest is  Ref.~\onlinecite{bell2012superinductor}, where the attainable ratio $\sqrt{E_C/E_L}$ is predicted to surpass the resistance quantum by more than two orders of magnitude. Moreover, recent experiments show that embedding a circuit within a resonator or a 3D cavity increases its relaxation time due to screening of the vacuum impedance, making pure dephasing a significant source of decoherence. \cite{paik2011observation, rigetti2012superconducting} Also, dephasing due to measurement back-action cannot be overcome completely. \cite{blais2004cavity} The results presented here 
are relevant for these scenarios. 

In the  fluxonium set-up, the role of a 3D cavity would thus be primarily that of insulating the circuit from the vacuum modes, so that the relaxation rate is reduced. In order to keep the device addressable, the circuit needs to be have significantly larger electrodes, which will decrease its capacitive energy. As we shall see, reducing the capacitive energy is detrimental for dephasing, but this can again be counteracted by increasing the value of the inductance.~\cite{masluk2012microwave, bell2012superinductor} 

\section{Decoherence and the \\
effective capacitance description}

Decoherence due to coupling of the qubit to the internal degrees of freedom of the superconducting material thus still poses a problem. \cite{catelani2011relaxation, catelani2012decoherence} To account for decoherence in the fluxonium circuit, we model its coupling to an environment using a \emph{Caldeira-Leggett}-type bath of harmonic oscillators. \cite{devoret1995quantum} The total 
Hamiltonian is
\bqa
\mathcal{H} &=& E_Cn^2 + E_L\phi^2 - E_J\cos(\phi-\theta) \nonumber\\
  &&+ \sum_q \hbar\omega_q b^\dagger_q b_q
  -\phi\sum_q\hbar g_q(b^\dagger_q + b_q) ,
  \label{fluxonium}
\eqa
where $E_C = \frac{(2e)^2}{2C}$ is the capacitive energy of the circuit, $E_L = \frac{(\Phi_0/2\pi)^2}{2L}$ is the energy associated with the linear inductive element, and $E_J = \frac{I_c\Phi_0}{2\pi}$ is the Josephson energy. The superconducting phase across the Josephson junction, $\phi$, and the number of Cooper pairs stored in the island, $n$, are quantised such that $[\phi,n] = i$, while $\theta = \frac{2\pi\Theta}{\Phi_0}$ is the dimensionless flux threading through the circuit. The first line in Eq.~(\ref{fluxonium}) corresponds to the fluxonium Hamiltonian, whose numerical diagonalization without bath coupling is depicted in Fig.~\ref{fig1}(b). The second line gives the environment and interaction Hamiltonians, 
from which a master equation is derived. Here, $b_q^{\dagger}$ ($b_q$) is the creation (annihilation) operator for environmental mode $q$, with frequency $\omega_q$ and system-bath coupling constant $g_q$. 

Varying the ratio $E_J/E_C$, using for instance a SQUID-like Josephson junction, dramatically changes the spectrum of fluxonium, and allows insight into the conditions under which a suppression of pure dephasing is expected. For values of $E_J/E_C\sim1$ and smaller, phase slips are common and the superconducting phase does not have a well-defined value. The current flowing through the circuit is negligible - 
it is in an insulating phase. 
In this regime, the lower-lying eigenstates of the circuit can be well approximated by harmonic oscillator states with an effective transition frequency $\omega^* = \frac{2\sqrt{E_LE_{C^*}}}{\hbar}$, where $E_{C^*} = \frac{(2e)^2}{2C^*}$, and the effective capacitance $C^*$ can be defined as
\be
C^* = (2e)^2\left[\frac{d^2\epsilon(\tilde n)}{d\tilde n^2}\right]^{-1}_{\tilde n=0}.
\label{eff_cap}
\ee
Here, $\epsilon(\tilde n)$ is the charge dispersion relation, and $2e\tilde n$ is  
the charge 
analogy of the quasimomentum in a one-dimensional lattice.

\begin{figure}[!t]
\centering
\subfigure{
\includegraphics[scale=.17]{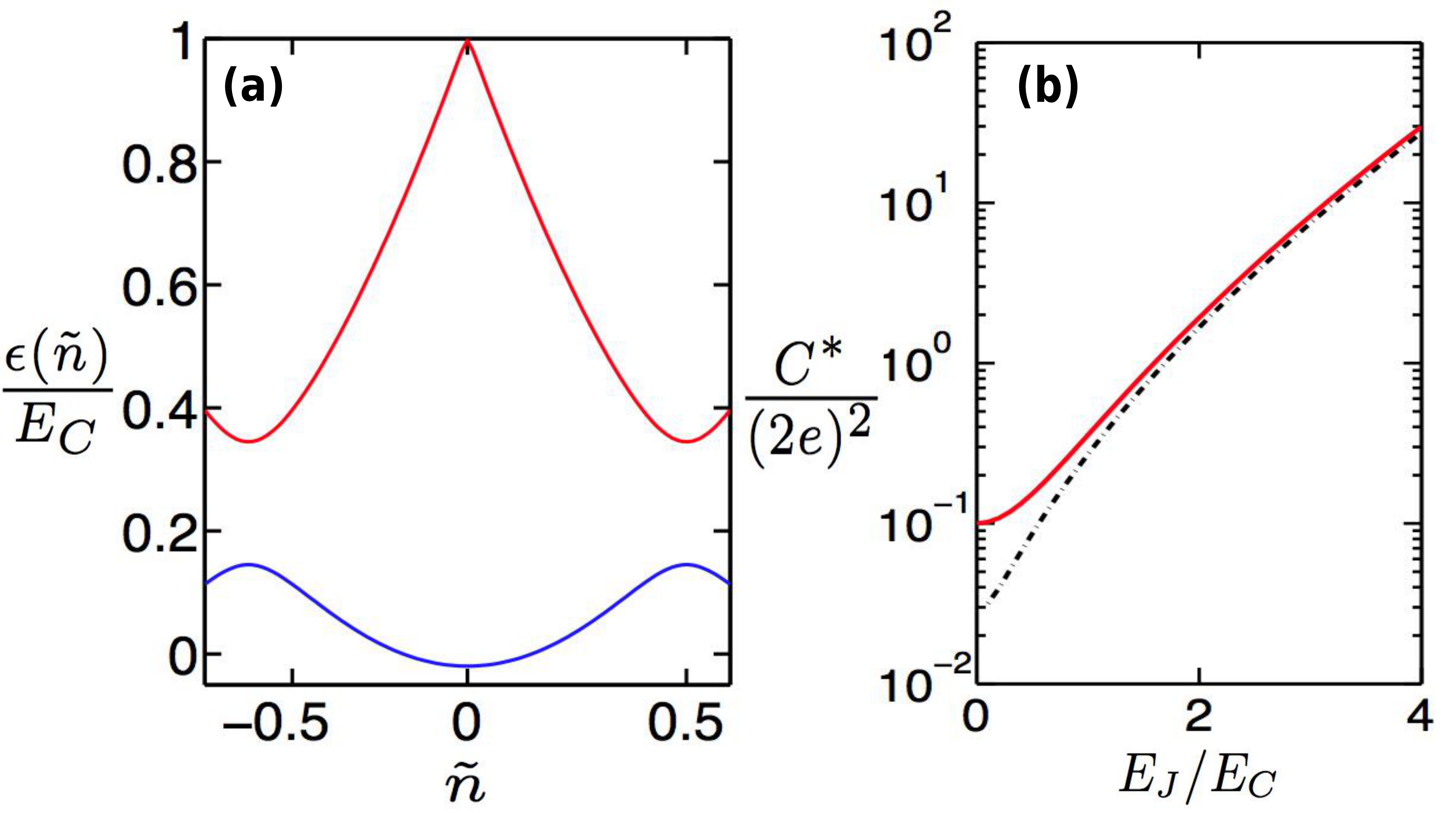}}
\subfigure{
\includegraphics[scale=.2]{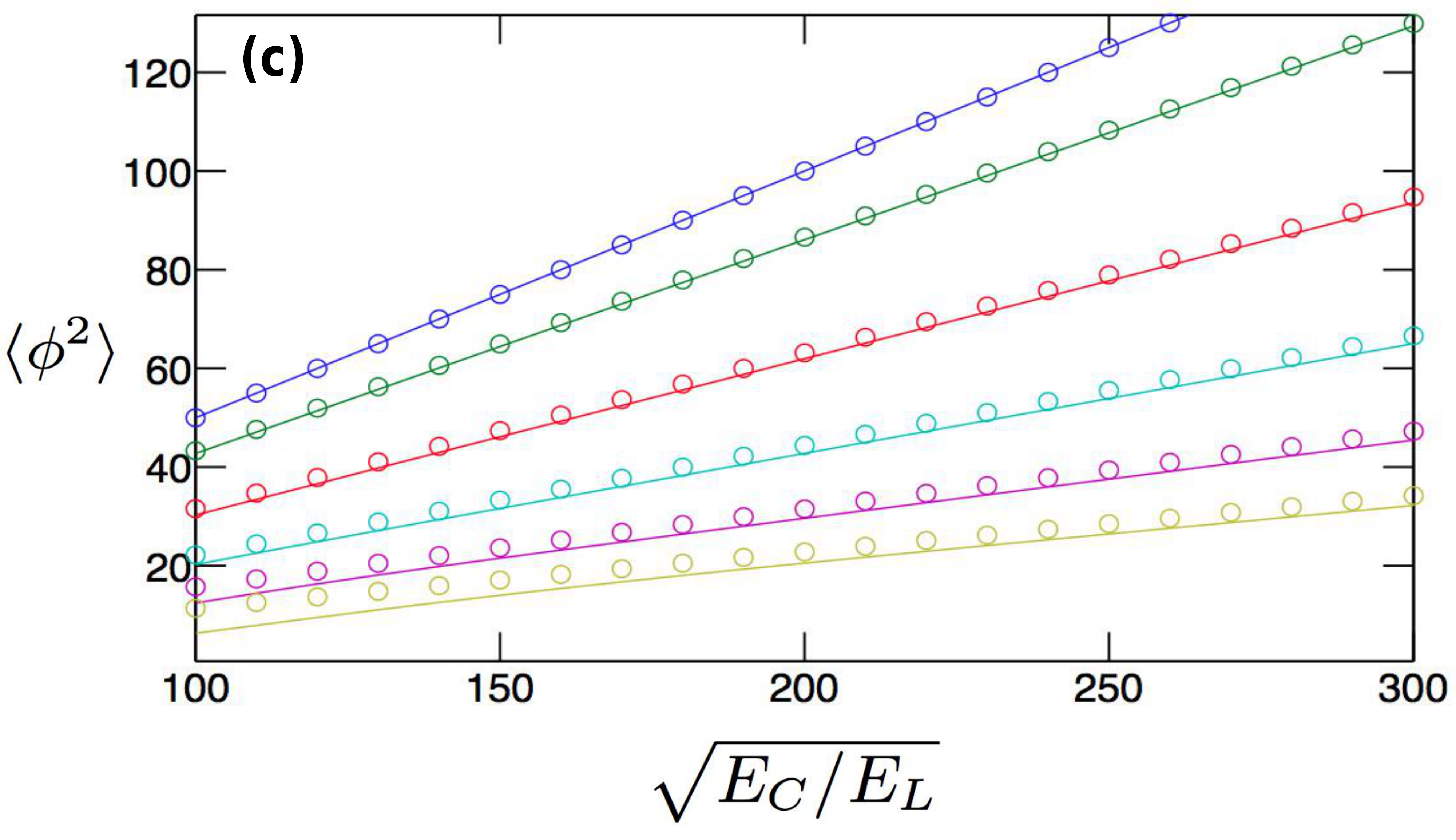}}
\caption{{(a)} Dispersion relation $\epsilon(\tilde n)$ of the circuit for the ground state (blue, lower) and the first excited state (red, upper) for $E_J/E_C = 0.5$. Close to the $\tilde n = 0$ the ground state energy is almost parabolic with a modified curvature, caused by the presence of an anticrossing at $\tilde n = \pm 0.5$. {(b)} Effective capacitance $C^*/(2e)^2$, in units of the capacitive energy $E_C$. The red (solid) curve corresponds to $C^*$ computed numerically using Eq.~(\ref{eff_cap}), while the dash-dotted line is calculated from Eqs.~(\ref{qps_amplitude}) and~(\ref{cstar}). {(c)} Variance of the wavefunction at $\theta=0$ for a range of values, $E_J/E_C \in \{0,2/5,4/5,...,2\} $, where $E_J/E_C $ increases in steps of $2/5$ from top to bottom. The solid lines are obtained by numerically diagonalising the fluxonium circuit Hamiltonian, and the circles correspond to the variance derived using the effective capacitance picture. Note that the approximation gets better as $\sqrt{E_C/E_L}$ increases.}
\label{fig2}
\end{figure}

In fact, for $E_J/ E_C \sim 1$, the tight-binding approximation applies and the charge dispersion relation can be written as $\epsilon(\tilde n) = -2t\cos(2\pi\tilde n)$ [see Fig.~\ref{fig2}(a)]. The tunneling amplitude $t$ can be obtained via instanton methods, \cite{matveev2002persistent, koch2007charge} which with our energy definitions is given by the expression
\be
t = \frac{4(2E^3_JE_C)^{1/4}}{\sqrt{\pi}}\exp\left(-8\sqrt{\frac{E_J}{2E_C}}\right).
\label{qps_amplitude}
\ee
It is thus possible to derive an analytic expression for the effective capacitance in this regime, using Eq.~(\ref{qps_amplitude}) and the tight-binding approximation, to give 
\be
C^* = \frac{(2e)^2}{8t\pi^2},
\label{cstar}
\ee
which can in turn be used to obtain the variance of the wavefunctions through the usual harmonic oscillator relation $\sigma^2_k = \frac{2k+1}{2}\sqrt{\frac{E_{C^*}}{E_L}}$. Figs.~\ref{fig2}(b) and (c) depict the effective capacitance of the circuit computed both numerically and analytically, and the fluctuations of the phase as a function of the ratio $\sqrt{E_C/E_L}$, respectively. 

The effective capacitance description will break down, however, when the wavefunction in the superconducting phase basis does not expand over several minima, i.e., when the phase particle does not ``see" a crystal anymore. In the charge basis, this happens when the wavefunction is so wide that that the curvature of the dispersion relation at the origin $\tilde n=0$, which arises in the expression for the effective capacitance, cannot be evaluated. We expect this to happen when the wavefunction starts to become localized in the phase basis, so a persistent current can be measured.
 
\begin{figure}[!t]
\centering
\includegraphics[scale=.22]{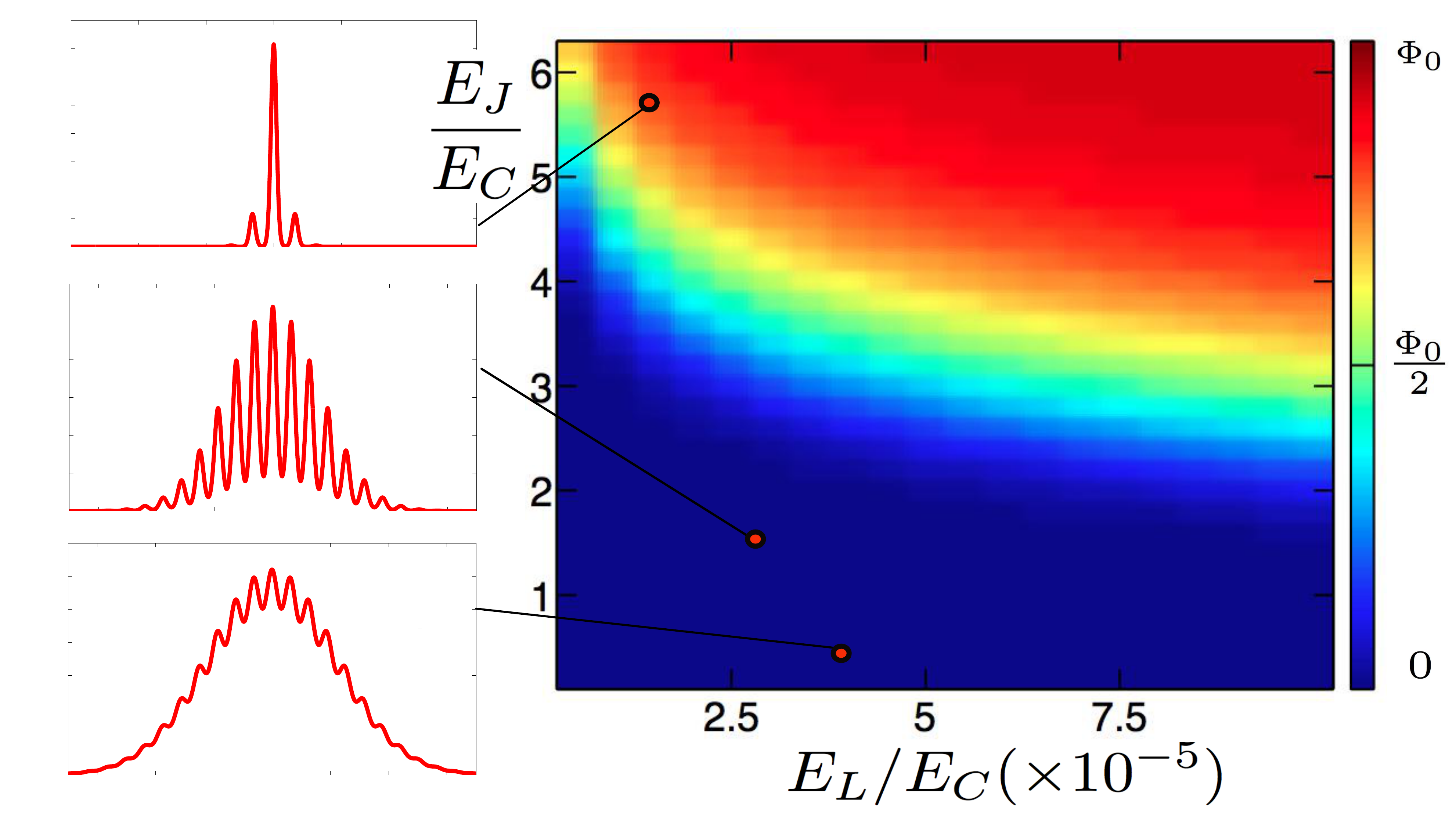}
\caption{Maximum value of the persistent current normalised by the inductive energy, $I_p (\times2L)$. The dark blue area at the bottom corresponds to the insulating regime. Increasing the ratio $E_J/E_C$ causes the circuit to enter a superconducting regime (red area at the top). This transition depends also on the ratio $E_L/E_C$, since smaller inductive energies result in a wider wavefunction, and the transition then takes place for larger values of the Josephson energy. The insets on the left illustrate the shape of the wavefunction at different points on the diagram. Whereas the phase is distributed roughly according to a Gaussian distribution in the insulating regime, it becomes localised in the superconducting regime.}
\label{fig3}
\end{figure}

As $E_J/E_C$ increases, phase slips are exponentially reduced [Eq.~(\ref{qps_amplitude})] and the phase does indeed become localized. In this regime a persistent current arises and the ground state strongly depends on the external flux $\theta$. The transition from the insulating to the superconducting regime is illustrated in Fig.~\ref{fig3}. It may in fact be possible to switch {\it in situ} between both regimes by substituting the Josephson junction in Fig.~\ref{fig1}(a) by a SQUID-like, flux dependent junction and varying the external flux. In the insulating regime, the wavefunctions spread over several minima of the periodic potential and fluctuations of the external flux, which cause the relative offset between the parabolic and periodic components of the potential to fluctuate, hardly influence the wavefunctions. In the superconducting regime, the wavefunctions are localized in a minimum of the periodic potential and are highly susceptible to fluctuations of the external flux. The results of Fig.~\ref{fig3}, and the preceding analysis, therefore suggests that in order to effectively suppress pure dephasing we should look at regimes of large $E_C/E_L$, as this will allow us to remain in the insulating regime for a larger range of $E_J/E_C$, thus minimising the adverse effect on the circuit anharmonicity.

\begin{figure}[!t]
\centering
\subfigure{
\includegraphics[scale=.205]{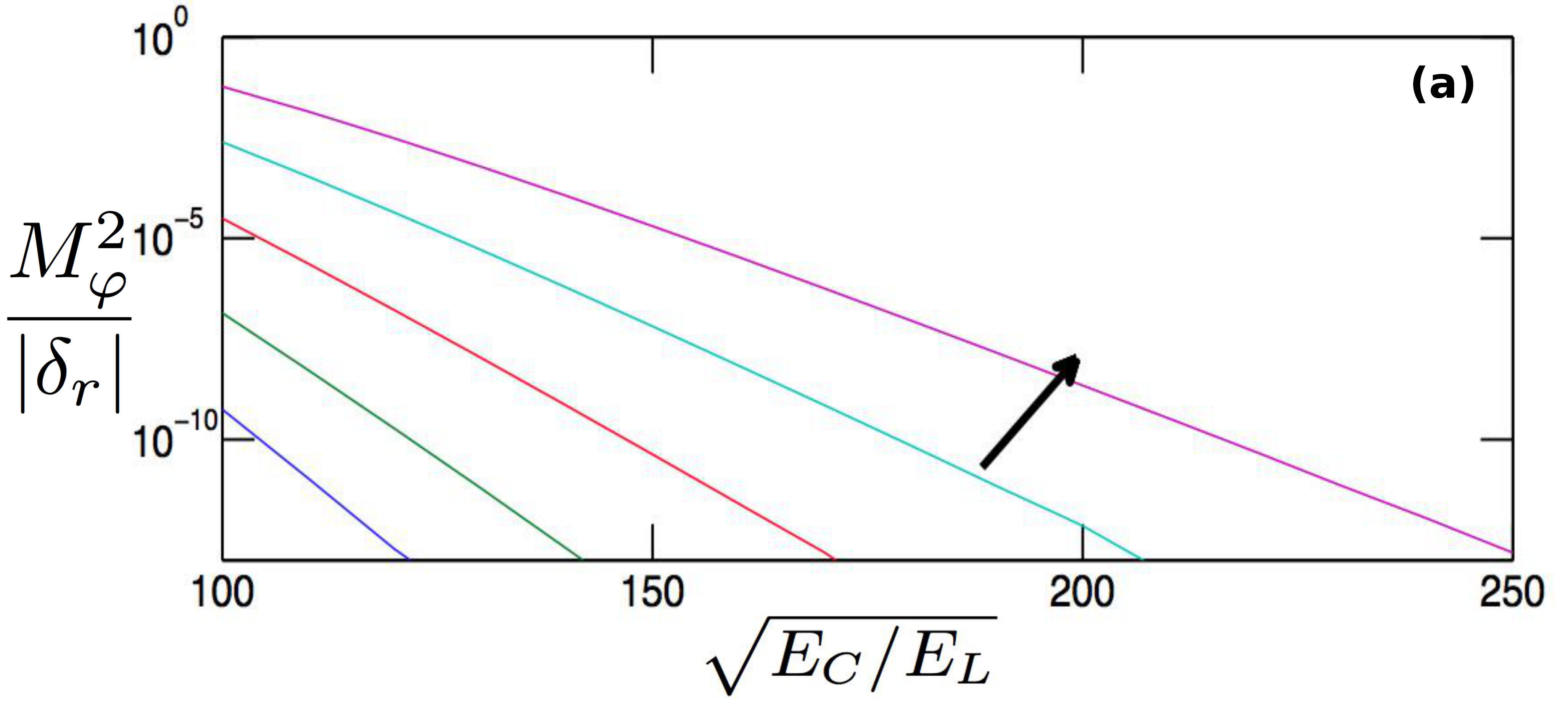}}
\subfigure{
\includegraphics[scale=.21]{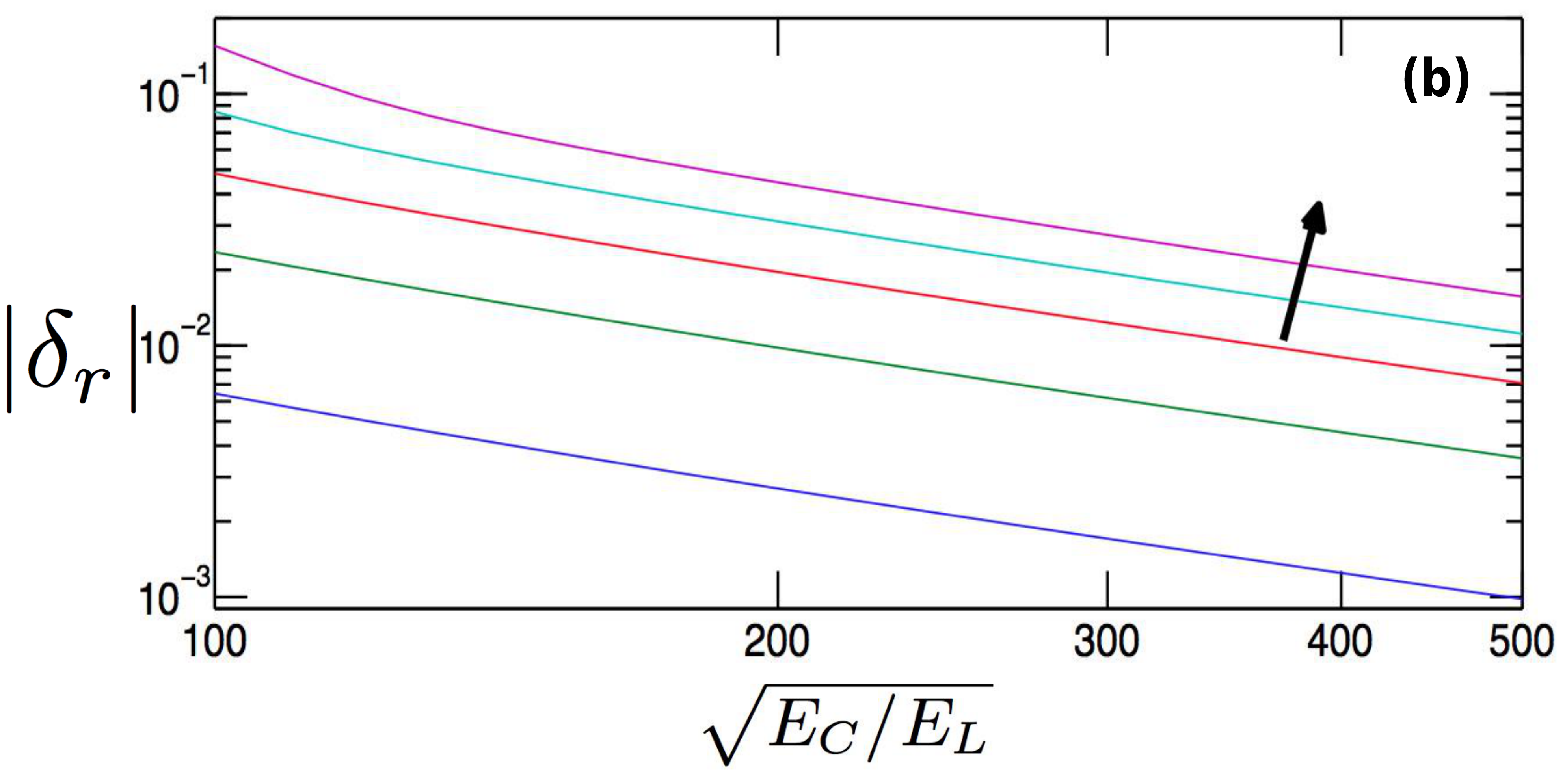}}
\caption{{(a)} The dephasing rates are proportional to $M^2_\varphi=|\langle 1|\phi|1\rangle - \langle 0|\phi|0\rangle|^2$. The ratio $M^2_\varphi/\delta_r$ vanishes as $\sqrt{E_C/E_L}$ increases. The different lines correspond to the values $E_J/E_C \in \{1/5,2/5,...,1\} $, where $E_J/E_C $ increases in steps of $1/5$ in the direction indicated by the arrow.  {(b)} With the same colour code as above ($E_J/E_C $ again increases in steps of $1/5$ in the direction indicated by the arrow), this \emph{log-log} plot shows that the relative anharmonicity $\delta_r$, which will ultimately determine the minimum length of the pulses, decays only algebraically. }
\label{fig4}
\end{figure}

In the fluxonium circuit eigenbasis (denoted $\ket{k}$, with energies $\epsilon_k$), the total Hamiltonian of Eq.~(\ref{fluxonium}) reads
\bqa
\mathcal{H} &=& \sum_k \epsilon_k \ketbra{k}{k} + \sum_i \hbar\omega_q b^\dagger_qb_q  \nonumber\\
&& + \sum_{k,k'}M_{kk'}\ketbra{k}{k'}\sum_q\hbar g_q(b^\dagger_q + b_q),
\eqa
where $M_{kk'}= \langle k|\phi|k'\rangle$. 
Writing a Markovian master equation for the circuit truncated to its two lowest-lying eigenstates allows us to identify the pure dephasing rate of the qubit:
 \be
\Gamma_\varphi = \frac{\pi M^2_\varphi}{4}\lim_{\omega\rightarrow 0}J(\omega)\coth\frac{\beta\hbar\omega}{2}, 
 \ee
 where $M^2_\varphi=|\langle 1|\phi|1\rangle - \langle 0|\phi|0\rangle|^2$, $J(\omega)=\sum_q|g_q|^2\delta(\omega-\omega_q)$ is the bath spectral density, and $\beta=(k_BT)^{-1}$ with $T$ the bath temperature and $k_B$ the Boltzmann constant. 
Our numerical simulations show that in the insulating regime $M_{\varphi}^2$, and hence pure dephasing, is in fact exponentially suppressed as $\sqrt{E_C/E_L}$ increases, as shown in Fig.~\ref{fig4}(a). This is consistent with protection against dephasing for low frequency noise. \cite{koch2009charging} In fact, assuming that the effective capacitance approximation holds so that the wavefunctions are approximately harmonic,  variations of the energy difference between the two lowest lying eigenstates can be written as:
\bqa
 \frac{d}{d\theta}\Delta_{10}&=& -E_J\frac{d}{d\theta} \left( \langle 1|\cos(\phi-\theta)|1\rangle 
- \langle 0|\cos(\phi-\theta)|0\rangle \right), \nonumber\\
&\approx& -E_J\sin(\theta)\sigma^2_0e^{-\frac{\sigma^2_0}{2}},
\label{variance}
\eqa
where $\sigma^2_0 = \sqrt{E_{C^*}/E_L}$.

Turning now to the effect on anharmonicity, a useful parameter to characterise the quality of the two-level system truncation is the relative anharmonicity, $\delta_r = \delta/\Delta_{10}$. 
As shown in Fig.~\ref{fig4}(b), our calculations predict that the relative anharmonicity decays only algebraically as the impedance increases in the considered regime, which is in stark contrast with the exponential reduction of the dephasing matrix element shown in Fig.~\ref{fig4}(a). As can be inferred from Eq.~(\ref{leakage}), in order to keep the leakage probability small, $p_L \ll 1$, the gate time $\tau$ must remain longer than the inverse anharmonicity, i.e., $\tau > \hbar/\delta$. Shortest pulse lengths currently achievable are of the order of 10 ns, \cite{paik2011observation} which remains longer than $\hbar/(\delta_r\Delta_{10})$ in the range of values of $\delta_r$ we obtained, and for typical transition frequencies of about $2\pi\times10$ GHz. \cite{manucharyan2012evidence}

Superinductors with $L\ge 10^3$~nH have been realised experimentally, and it is possible to create inductive elements in excess of  $L = 10^4$~nH.~\cite{bell2012superinductor} To estimate the potential reduction in the pure dephasing rate at such inductances we take experimental values from Ref.~\onlinecite{manucharyan2012evidence}, for which $\Gamma_\varphi \approx 400$~KHz at $\sqrt{E_C/E_L}\approx 2.6$ and $E_J/E_C\approx 2.5$, which we find corresponds to $M_{\varphi}^2\approx30$. Hence, taking $L = 10^4$~nH, a capacitive energy similar to experimental values in 3D cavities,~\cite{paik2011observation} that is $E_C/h \approx 300$~MHz, and for a Josephson energy of $E_J/h=150$~MHz, we find numerically that $M_{\varphi}^2$ can be reduced by almost an order of magnitude, corresponding to a pure dephasing rate of $\Gamma_\varphi \approx 50$~KHz.  
We note that our chosen value of $E_J$ implies a critical current of approximately $300 \textrm{~pA}$, which may be difficult to achieve using current state-of-the-art Josephson junctions. Recent advances \cite{meckbach2013submicron,wu2013fabrication} in nanoscale fabrication have led to junctions as small as $100\times 100 \textrm{~nm}^2$, while current densities as low as $20~\textrm{Acm}^{-2}$  have been measured,~\cite{weber2012aps} corresponding to a potential critical current of around $2000 \textrm{~pA}$.
Moreover, although it is clearly a technological challenge, the further reduction of junction size is still a very active area of research.
Indeed, although junction miniaturisation is generally pursued to enhance device performance for other reasons, our results highlight the additional benefit of fabricating sub-micrometre scale junctions to minimise dephasing. An alternative approach would be to use a SQUID-like junction, such that $E_J$ depends on an external control flux. Fluctuations of this control flux could then constitute an additional source of dephasing, but this should again be suppressed 
in the insulating regime.

\begin{figure}[!t]
\centering
\includegraphics[scale=.225]{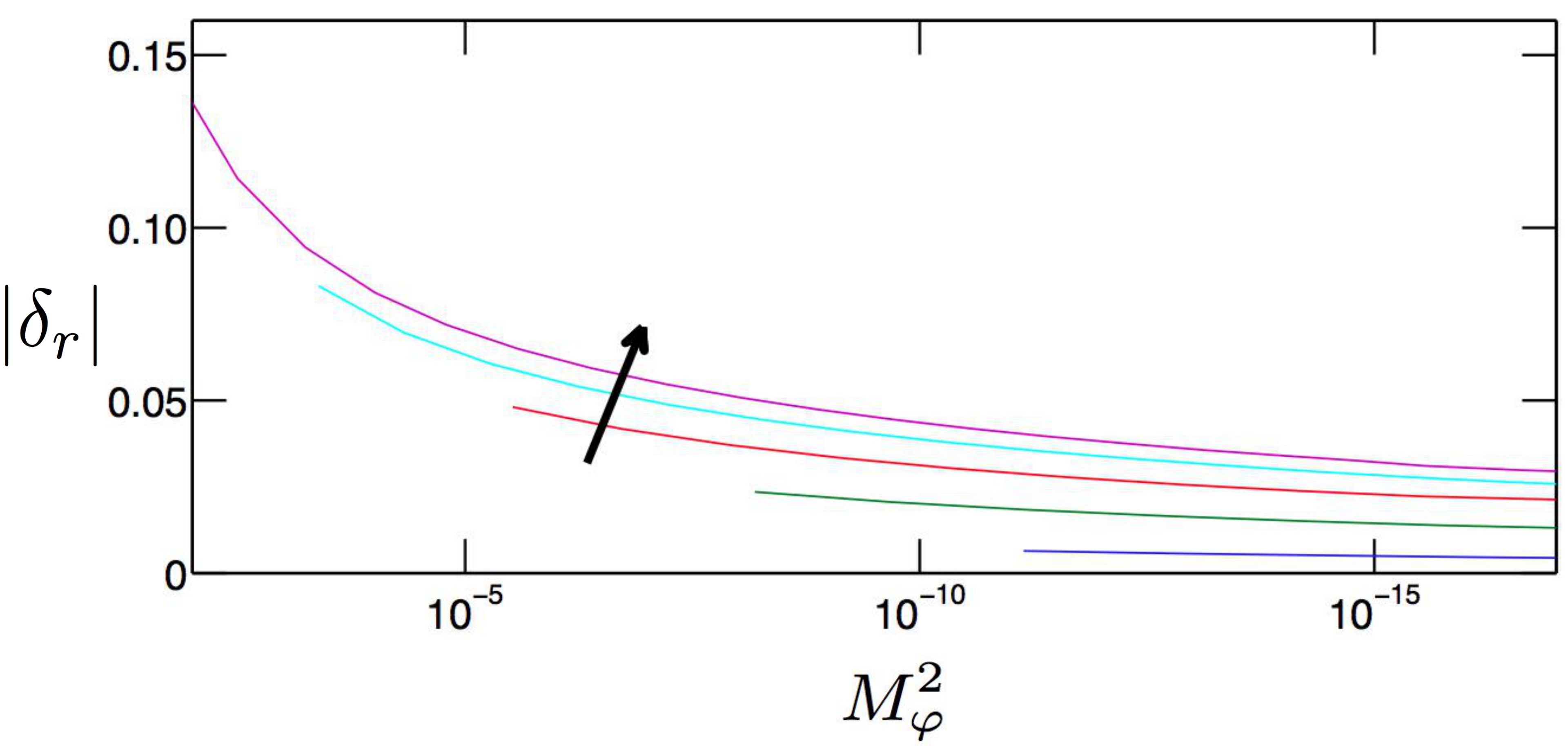}
\caption{Tradeoff between the pure dephasing matrix element $M^2_\varphi$ and the relative anharmonicity, obtained from combining 
Figs.~\ref{fig4}(a) and (b). The curves correspond to $E_J/E_C \in \{1/5,2/5,..., 1\} $, increasing in steps of $1/5$ in the direction indicated by the arrow. For $E_J/E_C = 2/5$ and above, the relative anharmonicity is large enough to ensure low leakage rates even at very high impedances.}
\label{fig5}
\end{figure}

Finally, in Fig.~\ref{fig5}, we show a tradeoff between the relative anharmonicity and the pure dephasing matrix element $M^2_\varphi$, related to leakage and dephasing errors, respectively. We observe that, in the insulating regime, it is possible to exponentially reduce the dephasing probability per gate without compromising significantly the tolerance to leakage errors. The threshold for loss-tolerant quantum computing is significantly higher than for computational errors,~\cite{stace2009thresholds, stace2010error, barrett2010fault,wang2011} and leakage out of the computational subspace is an inconvenience relatively easier to circumvent than detecting and correcting unknown computational errors. Whenever possible, it is thus desirable to trade computational errors for loss errors, which can be done by increasing the impedance of the circuit.

\section{Summary and Discussion}

We have shown that it is possible to exponentially suppress the pure dephasing rate of the fluxonium qubit  caused by coupling to a Markovian bath, at the expense of (slightly) reduced anharmonicity. The anharmonicity only decays algebraically, however, allowing for a tradeoff in which the length of the driving pulses need not be increased as significantly in order to compensate for leakage errors. Thus, both the dephasing and leakage error rates may be made small simultaneously.  At larger values of the circuit's impedance, the lowest-lying states are better approximated by harmonic oscillator states in the insulating regime, and the transition from the insulating to the superconducting phases occurs at increasing values of the ratio $E_J/E_C$. The independence of the harmonic oscillator states to external flux variations can thus be inherited by the fluxonium states, though without a complete loss of anharmonicity necessary to define a qubit. Interestingly, this is achieved in the same regime in which low-frequency external flux noise is suppressed,~\cite{koch2009charging} though here for a generic Markovian environment.  
In fact, this result should hold in the case of a non-Markovian environment too, since it relies on the suppression of the matrix elements $M_\varphi$, rather than on any particular form of noise. 

Experimental advances towards realisation of very high impedances,~\cite{bell2012superinductor} as well as towards to the inhibition of relaxation processes in 3D cavities, \cite{paik2011observation, rigetti2012superconducting} make our results relevant in the design of the next generation of superconducting qubits. In our case, the role of a 3D cavity would be mainly that of insulating the circuit from the vacuum modes, so that the relaxation rate is reduced. In order to maintain full control of the circuit larger capacitors are needed, which in turn results in a lower capacitive energy. Reducing the capacitive energy contributes to locking of the centre of the wavefunction to the value of the external flux, which is detrimental. However, increasing the value of the inductance contributes to spreading of the wavefunction, and thus it becomes less sensitive to external flux variations.

We 
emphasise that the fluxonium circuit offers protection against dephasing in new ways and that a substantial decrease of pure dephasing rates is already achievable with state-of-the-art devices.  As a final remark, we would also like to mention that this circuit is a relevant step towards the realisation of a topologically protected qubit based on a superconducting current mirror, in which the computational states are stored in a degenerate ground state, and a relevant source of decoherence is expected to be pure dephasing.~\cite{kitaev2006protected}

\acknowledgements

This work was completed shortly after the tragic death of one of the authors, Sean Barrett. It is with immense gratitude that we acknowledge his close friendship, his guidance and his insight, which we sorely miss. 
We would like to thank Tom Stace for critical reading of the manuscript and Andrew Doherty for comments. This work was supported by the EPSRC, Imperial College London, and the Royal Society.

\end{document}